\documentclass[aps,showkeys,showpacs,twocolumn]{revtex4}
\usepackage{amsmath}
\usepackage{amscd}
\usepackage{graphicx}
\usepackage{subfigure}
\usepackage{appendix}
\usepackage{amsfonts}
\usepackage{multirow}
\usepackage{bm}
\usepackage{float}

\begin{document}

\title[Short Title]{Coherent control in quantum open systems: An approach for accelerating dissipation-based quantum state generation}

\author{Ye-Hong Chen$^{1,2}$}
\author{Zhi-Cheng Shi$^{1,2}$}
\author{Jie Song$^{3}$}
\author{Yan Xia$^{1,2,}$\footnote{E-mail: xia-208@163.com}}
\author{Shi-Biao Zheng$^{1,2}$}

\affiliation{$^{1}$Department of Physics, Fuzhou University, Fuzhou 350116, China\\
             $^{2}$Fujian Key Laboratory of Quantum Information and Quantum Optics (Fuzhou University), Fuzhou 350116, China\\
             $^{3}$Department of Physics, Harbin Institute of Technology, Harbin 150001, China}


\begin{abstract}
  In this paper, we propose an approach to accelerate the dissipation dynamics for quantum state generation {with Lyapunov control}.
  The strategy is to add target-state-related coherent control fields
  into the dissipation process to intuitively improve the evolution speed. By applying the current approach,
  without losing the advantages of dissipation dynamics, the target stationary states can be generated in a much shorter time
  as compared to that via traditional dissipation dynamics.
  As a result, the current approach containing the advantages of coherent unitary dynamics and dissipation dynamics
  allows for significant improvement in quantum state generation.
\end{abstract}


\maketitle
For years, quantum dissipation has been treated as a resource
rather than as a detrimental effect to generate a quantum state
\cite{Prl106090502,arXiv10052114,Njp11083008,Nat4531008,Jpa41065201,Prl107120502,Np5633} in quantum open systems
modeled by the Lindblad-Markovian
master equation \cite{Cmp48119} ($\hbar=1$)
\begin{align}\label{eq0-1}
  \dot{\rho}=&-i[H_{0},\rho]+\mathcal{L}\rho,\cr
  \mathcal{L}\rho=&\sum_{k}L_{k}\rho L_{k}^{\dag}-\frac{1}{2}(L_{k}^{\dag}L_{k}\rho+\rho L_{k}^{\dag}L_{k}),
\end{align}
where the overdot stands for a time derivative and $L_{k}$ are the so-called Lindblad operators.
By using dissipation, one can generate high-fidelity quantum states without
accurately controlling the initial state or the operation time
(usually, the longer the operation time is, the higher is the fidelity). Besides, dissipation dynamics is
shown to be robust against parameter (instantaneous) fluctuations \cite{Prl106090502}.
Due to these advantages, many schemes
\cite{arXiv11101024,Pra84022316,Pra83042329,Pra82054103,Prl89277901,Prl117210502,Pra84064302,Nat504415,Prl111033607,Pra95022317,Npto10303,Prl117040501,Prl115200502}
have been proposed for dissipation-based quantum state generation in recent years based on different physical systems.

Generally speaking, to generate quantum states by quantum dissipation, the key point is to find (or design)
a unique stationary state (marked as $|S\rangle$) which can not be transferred to other states
while other states can be transferred to it. 
That is, the reduced system should satisfy
\begin{align}\label{eq0-3}
  H_{0}|M\rangle\neq 0,\ H_{0}|S\rangle=0,\   \tilde{L}_{k}^{\dag}|S\rangle\neq 0,\ \tilde{L}_{k}|S\rangle=0,
\end{align}
where $|M\rangle$ ($M\neq S$) are the orthogonal partners of the state $|S\rangle$ in a reduced system satisfying
$\langle M|S\rangle=0$ and $\sum_{M}|M\rangle\langle M|+|S\rangle\langle S|=\bm{1}$,
and $\tilde{L}_{k}$ are the effective Lindblad operators.
Hence, if the system is in $|M\rangle$, it will always be transferred to other states because
$H_{0}|M\rangle\neq 0$ and $\tilde{L}_{k}^{\dag}|S\rangle\neq 0$, while if the system is in $|S\rangle$, it remains invariant.
Therefore, the process of pumping and decaying 
continues until the system is finally stabilized into the stationary state $|S\rangle$.

To show such a dissipation process in more detail,
we introduce a function $\dot{V}$ to describe the system evolution speed,
where $V=\text{Tr}(\rho\rho_{s})$ {is known as the Lyapunov
function \cite{Ddbook}} and $\rho_{s}$ is the density matrix of the target state $|S\rangle$.
{Lyapunov control is a form of local optimal control with numerous variants \cite{Ddbook,Pra80052316,A4498,Njp11105034},
which has the advantage of being sufficiently simple to be amenable to rigorous analysis and has been
used to manipulate open quantum systems \cite{Pra80052316,Pra80042305,Pra82034308}. For example, Yi \emph{et al.} proposed a scheme in 2009 to drive a finite-dimensional quantum system into
the decoherence-free subspaces by Lyapunov control \cite{Pra80052316}.}

When the system evolves into a target state at a final time $t_{f}$, i.e., $\rho|_{t=t_{f}}\rightarrow\rho_{s}$,
$V$ approaches a maximum value $V=1$.
Based on Eqs. (\ref{eq0-1}) and (\ref{eq0-3}), we find
\begin{eqnarray}\label{eq0-4}
  \dot{V}=\text{Tr}[(-i[H_{0},\rho]+\mathcal{L}\rho)\rho_{s}]
         =\sum_{k}\Gamma_{k}\langle E_{k}|\rho| E_{k}\rangle\geq0,
\end{eqnarray}
in which we have assumed $\tilde{L}_{k}=\sqrt{\Gamma_{k}}|S\rangle\langle E_{k}|$,
with $\Gamma_{k}$ being the effective dissipation rates and $|E_{k}\rangle$ being the effective excited states.
Obviously, the evolution speed strongly dependents on the effective dissipation rates and the total population of effective excited states.
Hence, according to the dissipation dynamics,
we have $\langle E_{k}|\rho|E_{k}\rangle\rightarrow 0$ when $t\rightarrow\infty$, which means $\dot{V}|_{t\rightarrow \infty}=0$.

However, as is known, such a process is generally much slower than a unitary evolution process because of the small effective dissipation rates.
It would be a serious issue to realize large-scale integrated computation if it
takes too long to generate the desired quantum states.
In view of this, the preponderance of dissipation-based approaches would lose
if a future technique would present an ideal dissipation-free system.
Therefore, accelerating the dissipation dynamics without losing its advantages should signal a significant improvement for quantum computation.
Now that a unitary evolution process is much faster than a dissipation process,
we are guided to ask, is it possible to accelerate the dissipation dynamics by using coherent control fields?
{In Ref. \cite{Pra80052316} the authors
mentioned that Lyapunov control may have the ability to shorten the convergence time for an open system.
Therefore, in this paper, we will seek additional coherent control fields according to Lyapunov control
to accelerate dissipation dynamics.}

The strategy of accelerating dissipation dynamics is to add a simple and realizable coherent control Hamiltonian $H_{c}$
to increase the value of $\dot{V}$ in Eq. (\ref{eq0-4}).
The state evolution equation in this case becomes
\begin{align}\label{eq0-5}
  \dot{\rho}=-i[H_{0}+H_{c},\rho]+\mathcal{L}\rho,
\end{align}
where $H_{c}=\sum_{n}f_{n}(t)H_{n}$ is the additional control Hamiltonian,
$H_{n}$ are time independent, and control functions $f_{n}(t)$ are realizable and real valued.
The corresponding evolution speed reads
\begin{align}\label{eq0-6}
  \dot{V}_{a}=&\text{Tr}[(-i[H_{0},\rho]+\mathcal{L}\rho)\rho_{s}]\cr
                  &+\sum_{n}f_{n}(t)\text{Tr}[(-i[H_{n},\rho])\rho_{s}].
\end{align}
We use the symbol $\dot{V}_{a}$ to distinguish from the original evolution speed $\dot{V}$.
The control functions $f_{n}(t)$ should be carefully chosen to ensure that
$V_{a}|_{t=t_{f}'}=1$ and $\dot{V}_{a}|_{t=t_{f}'}=0$.
For this goal, the simplest choice for $f_{n}(t)$ is \cite{Pra80052316}
\begin{align}\label{eq0-7}
  f_{n}(t)=\text{Tr}[(-i[H_{n},\rho])\rho_{s}].
\end{align}
As can be seen from Eq. (\ref{eq0-3}), the Hamiltonian $H_{0}$ is just used
to ensure that $|S\rangle$ is a stationary state,
while, by adding additional coherent fields, it is easy to
find $(H_{0}+H_{c})|S\rangle\neq 0$ (for $\rho\neq\rho_{s}$ corresponding to $t< t_{f}$), which means
$|S\rangle$ is actually not a stationary state when $t<t_{f}$.
For $t\rightarrow t_{f}$, according to Eq. (\ref{eq0-7}), we have $f_{n}(t_{f})=0$ since $\rho|_{t=t_{f}}\rightarrow \rho_{s}$.
Thus, $H_{c}=0$, so that $|S\rangle$ becomes a unique stationary state when $t=t_{f}$.
That is, when $t<t_{f}$, the coherent fields and dissipation work together to
drive the system to state $|S\rangle$, while when $t\rightarrow t_{f}$,
the additional coherent fields vanish and the system becomes steady.
It can also be understood as, in the current approach, $|S\rangle$
is not a stationary state until the population is totally transferred to it.
Obviously, such a process is significantly different from the previous dissipation-based
schemes \cite{Prl106090502,Prl117210502,Pra84064302,Nat504415,Prl111033607,Pra95022317},
in which $|S\rangle$ is the unique stationary state during the whole evolution.

Usually, part of $H_{0}$ can be chosen as $H_{n}$ to make sure that $H_{n}$ is realizable.
In this case, the additional coherent control fields can be actually regarded as a
modification on Hamiltonian $H_{0}$. So, the current approach can be actually understood as
a parameter optimization approach for dissipation-based quantum state generation.
In the following, we will verify the accelerating approach with applications to quantum state generation.

\textit{Application I: Single-atom superposition state.} We first consider a three-level $\Lambda$ atom
with an excited state $|e\rangle$ and two ground states $|g_{1}\rangle$ and $|g_{2}\rangle$ to illustrate our accelerating approach.
The transition $|e\rangle\leftrightarrow|g_{1,(2)}\rangle$
is resonantly driven by a laser field with a Rabi frequency $\Omega_{1,(2)}$.
The Hamiltonian in the interaction picture is thus written as
$H_{0}=\Omega_{0}(\sin{\theta}|e\rangle\langle g_{1}|+\cos{\theta}|e\rangle\langle g_{2}|)+H.c.$,
where $\Omega_{0}=\sqrt{\Omega_{1}^{2}+\Omega_{2}^{2}}$ and $\theta=\arctan{\frac{\Omega_{1}}{\Omega_{2}}}$.
The Lindbald operators in this $\Lambda$ system associated with
atomic spontaneous emission are $L_{1}=\sqrt{\gamma_{1}/2}|g_{1}\rangle\langle e|$ and
$L_{2}=\sqrt{\gamma_{2}/2}|g_{2}\rangle\langle e|$, respectively.
Then, we introduce the orthogonal states $|S\rangle=\cos{\varphi}|g_{1}\rangle-\sin{\varphi}|g_{2}\rangle$ and
$|T\rangle=\sin{\varphi}|g_{1}\rangle+\cos{\varphi}|g_{2}\rangle$ to rewrite
the Hamiltonian $H_{0}$ as
  $H_{0}=\Omega_{S}|e\rangle\langle S|+\Omega_{T}|e\rangle\langle T|+H.c.$,
where $\Omega_{S}=\Omega_{0}\sin{(\theta-\varphi)}$ and $\Omega_{T}=\Omega_{0}\cos{(\theta-\varphi)}$.
Accordingly, by choosing $\gamma_{1}=\gamma_{2}=\gamma$, we obtain two effective Lindblad operators
$\tilde{L}_{S}=\sqrt{\gamma/2}|S\rangle\langle e|$ and $\tilde{L}_{T}=\sqrt{\gamma/2}|T\rangle\langle e|$.
It is clear that if we choose $\theta=\varphi$, the effective driving field between $|e\rangle$ and $|S\rangle$
with a Rabi frequency $\Omega_{S}$ will be switched off and the condition in Eq. (\ref{eq0-3}) will be satisfied.
In this case, according to dissipation dynamics, the system will be stabilized into the stationary state $|S\rangle$.
Beware that the present application example maybe similar with that in Ref. \cite{Pra82034308} proposed by Wang \emph{et al.} which used Lyapunov
control to drive an open system (with a four-level atom driven by two lasers) into a decoherence-free subspace.
Here we need to emphasize that, in this paper, we focus on analyzing the evolution speed and how the Lyapunov
control can accelerate the dissipation dynamics.

\begin{figure}[b]
 \scalebox{0.23}{\includegraphics {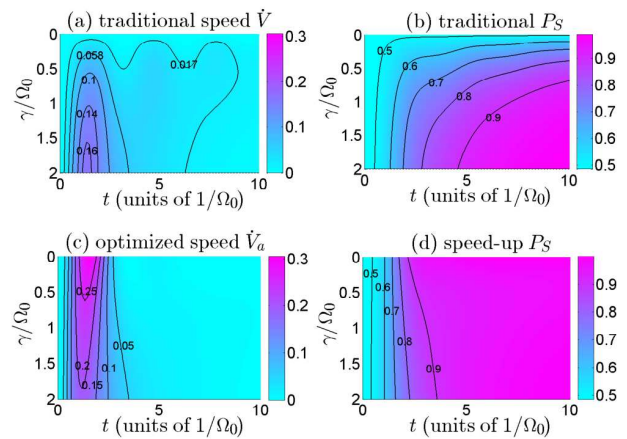}}
 \caption{
         Single-atom superposition state preparation:
         The comparison with respect to the evolution speed between the traditional
         dissipation dynamics and the accelerated dissipation dynamics.
         (a) and (c): The evolution speeds given according to Eq. (\ref{eq0-4}) and Eq. (\ref{eq0-6}) versus $\gamma$, respectively.
         (b) and (d): The time-dependent populations governed by the traditional dissipation dynamics and the accelerated dissipation dynamics, respectively.
         }
 \label{fig3}
\end{figure}

\begin{figure}[b]
 \scalebox{0.32}{\includegraphics {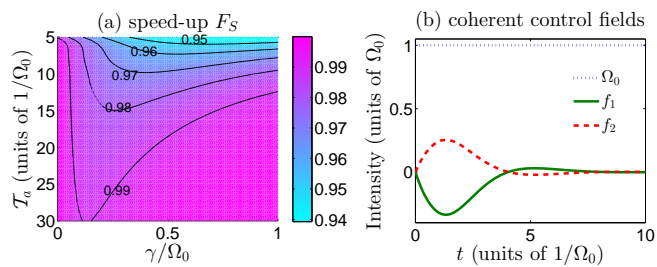}}
 \caption{
          Single-atom superposition state preparation:
          (a) Fidelity $F_{S}$ of the accelerated dissipation scheme versus $\mathcal{T}_{a}$ and $\gamma$,
          where the fidelity $F_{S}$ is defined by $F_{S}=\langle S|\rho|S\rangle|_{t=t_{f}}$ expressing the final population for the target state.
          (b) The coherent control fields for the accelerated dissipation scheme when $\gamma=0.8\Omega_{0}$.
          In general, the intensity for the additional coherent control fields should be smaller than
          $\Omega_{1,(2)}$.
         }
 \label{fig4}
\end{figure}

By choosing $t_{f}=10/\Omega_{0}$, the evolution speed $\dot{V}$ and time-dependent population for state $|S\rangle$ versus $\gamma$ are displayed in Figs. \ref{fig3} (a) and (b), respectively.
As shown in the figure, to obtain the target state $|S\rangle$ in a relatively high fidelity $\geq0.95$
within a fixed evolution time $t_{f}=10/\Omega_{0}$, the decay rate should be at least $\gamma\geq \Omega_{0}$
($P_{S}|_{t=t_{f}}=0.9506$ when $\gamma=\Omega_{0}$).
To accelerate such a process by additional coherent control fields,
we choose the control Hamiltonians $H_{n}$ as
  $H_{1}=\mu_{1}|e\rangle\langle g_{1}|+H.c.$ and
  $H_{2}=\mu_{2}|e\rangle\langle g_{2}|+H.c.$,
where $\mu_{1}$ and $\mu_{2}$ are two arbitrary time-independent
parameters used to control the intensities of the control fields.
By choosing $\mu_{1}=0.8$ and $\mu_{2}=0.6$ as an example,
the optimized evolution speed $\dot{V_{a}}$ given according to Eq. (\ref{eq0-6})
is shown in Fig. \ref{fig3} (c). Contrasting Figs. \ref{fig3} (c) with (a),
it is clear that the evolution speed has been significantly improved, especially,
when the decay rate $\gamma$ is relatively small.
For example, when $\gamma=0.5\Omega_{0}$, the maximum value of the evolution speed has been increased
from $\dot{V}^{max}\approx0.08$ to $\dot{V_{a}}^{max}\approx 0.26$.
While for a relatively large decay rate, the increasing effect is relatively weak.
This is because the control functions
$f_{n}(t)$ are mainly decided by the instantaneous distance $d=1-\text{Tr}(\rho\rho_{s})$ from the target state according to Eq. (\ref{eq0-7}).
In general, $f_{n}(t)$ are in direct proportion to $d$.
In a certain period of time,
more population will be transferred to the stationary state $|S\rangle$ with a relatively large decay rate (see Fig. \ref{fig3}).
That is, the instantaneous value of $\text{Tr}(\rho\rho_{s})$ decreases with the increase of $\gamma$.
Accordingly, the control functions $f_{n}(t)$ will fade away along with the increase of the decay rate $\gamma$.
To show the fidelity of the accelerated state generation in more detail, we display the fidelity
of the target state $|S\rangle$ versus operation time $\mathcal{T}=t_{f}-t_{i}$ ($t_{i}=0$ is the initial time) and decay rate $\gamma$ in Fig. \ref{fig4} (a).
For clarity, in the following, we will use the symbols $\mathcal{T}_{t}$ and $\mathcal{T}_{a}$ to express operation
times via traditional dissipation dynamics and accelerated dissipation dynamics, respectively.
It is clear from Fig. \ref{fig4} (a) that the efficiency of state generation has been
remarkably improved since a relatively high fidelity ($F_{S}\approx0.95$) of
the target state $|S\rangle$ can be achieved even when the operation time is only $\mathcal{T}_{a}=5/\Omega_{0}$.
The shapes for the additional control fields are shown to be smooth curves [see Fig. \ref{fig4} (b) with $\gamma=0.8\Omega_{0}$ as an example]
which can be easily realized in practice.
For example, one can use electrooptic modulators to implement such coherent fields.

Affected by the real experimental environment, 
there is usually a stochastic kind of noise that should be considered in realizing the scheme.
Assume that the Hamiltonian $H_{0}$ is perturbed by
some stochastic part $\eta H_{s}$ describing amplitude noise. A stochastic
Schr\"{o}dinger equation in a closed system (in the Stratonovich sense) is then
$\dot{\psi}(t)=[H_{0}+\eta H_{s}\xi(t)]\psi(t)$,
where $\xi(t)=\partial_{t}W_{t}$ is heuristically the time derivative of the Brownian motion $W_{t}$.
$\xi(t)$ satisfies $\langle\xi(t)\rangle=0$ and $\langle\xi(t)\xi(t')\rangle=\delta(t-t')$ because the noise should have zero mean and the noise
at different times should be uncorrelated. Then, we define $\rho_{\xi}(t)=|\psi_{\xi}(t)\rangle\langle\psi_{\xi}(t)|$, and
the dynamical equation without dissipation for $\rho_{\xi}$ is thus given as
\begin{align}\label{eq1-12}
  \dot{\rho}_{\xi}=-i[H_{0},\rho_{\xi}]-{i\eta}[H_{s},\xi\rho_{\xi}].
\end{align}
After averaging over the noise, Eq. (\ref{eq1-12}) becomes
  $\dot{\rho}\simeq-i[H_{0},\rho]-{i\eta}[H_{s},\langle\xi\rho_{\xi}\rangle]$,
where $\rho=\langle\rho_{\xi}\rangle$  \cite{Njp14093040}. According to Novikov's theorem in the case of white noise,
we have $\langle\xi\rho_{\xi}\rangle=\frac{1}{2}\langle\frac{\delta\rho_{\xi}}{\delta\xi(t')}\rangle|_{t'=t}=-\frac{i\eta}{2}[H_{s},\rho]$.
Hence, when both the amplitude noise and dissipation are considered, the dynamics of the open system will be governed by
\begin{align}\label{eq1-14}
  \dot{\rho}\simeq-i[H_{0},\rho]+\mathcal{N}\rho+\mathcal{L}\rho,
\end{align}
where $\mathcal{N}\rho=-\frac{\eta^2}{2}[H_{s},[H_{s},\rho]]$.

\begin{figure}[t]
 \scalebox{0.32}{\includegraphics {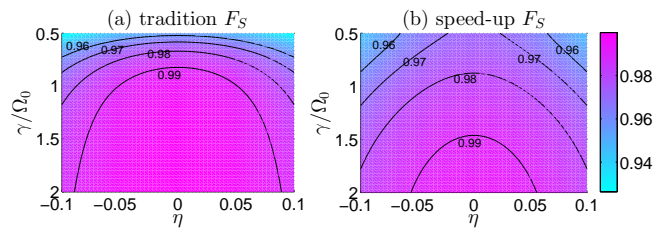}}
 \caption{
         Single-atom superposition state preparation:
         The comparison with respect to the robustness against amplitude-noise error between the traditional dissipation dynamics and the accelerated dissipation dynamics.
         (a) $F_{S}$ versus $\eta$ and $\gamma$ via traditional dissipation dynamics with $\mathcal{T}_{t}=20/\Omega_{0}$.
         (b) $F_{S}$ versus $\eta$ and $\gamma$ via the accelerated dissipation dynamics with $\mathcal{T}_{a}=10/\Omega_{0}$.
         Here $\mathcal{T}_{a}=10/\Omega_{0}$ is chosen to make the highest fidelity for each $\gamma$ in Fig. \ref{fig5} (b) to be
         the same as that in Fig. \ref{fig5} (a) as far as possible.
         }
 \label{fig5}
\end{figure}

For the current three-level scheme, we consider an independent amplitude noise in $\Omega_{1}$ as well as
in $\Omega_{2}$ with the same intensity $\eta^2$, and the noise term in Eq. (\ref{eq1-14}) is thus
\begin{eqnarray}\label{eq1-15}
  \mathcal{N}\rho=-\frac{\eta^2}{2}([H_{s1},[H_{s1},\rho]]+[H_{s2},[H_{s2},\rho]]),
\end{eqnarray}
where $H_{s1}=\Omega_{1}|e\rangle\langle g_{1}|+H.c.$ and $H_{s2}=\Omega_{2}|e\rangle\langle g_{2}|+H.c.$.
According to Eq. (\ref{eq1-14}), the robustness against
amplitude-noise error for the dissipation-based state generation without
the additional coherent control fields is shown in Fig. \ref{fig5} (a),
in which the operation time is chosen as $\mathcal{T}_{t}=20/\Omega_{0}$.
Only a $\sim2\%$ deviation will occur in the fidelity as shown in the figure with a relatively small decay rate $\gamma\leq1$
and the noise intensity is $\eta=0.1$. The robustness of the scheme against amplitude-noise error
is better when the decay rate gets larger.
For comparison, the robustness against the amplitude-noise error of the accelerated dynamics governed by
  $\dot{\rho}\simeq-i[H_{0}+H_{c},\rho]+\mathcal{N}\rho+\mathcal{L}{\rho}$,
is shown in Fig. \ref{fig5} (b) with operation time $\mathcal{T}_{a}=10/\Omega_{0}$.
The result shows the robustness of the accelerated scheme
with respect to amplitude-noise error is almost the same with that of the traditional scheme.
A stochastic noise with intensity $\eta=0.1$ also causes a deviation of about $2\%$ on the fidelity
when $\gamma\leq1$, and the influence of noise decreases with increasing $\gamma$.
That is, we have confirmed that the approach by adding coherent control fields can
realize the goal of accelerating the dissipation process without losing the advantage of robustness
against parameter fluctuations.

\begin{figure}[t]
 \scalebox{0.16}{\includegraphics {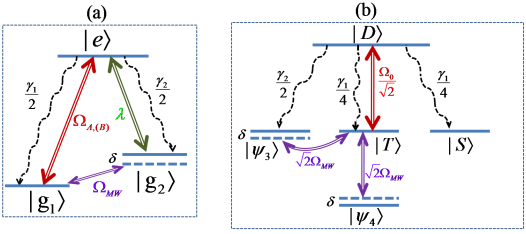}}
 \caption{
         Two-atom entanglement preparation:
         (a) Level diagram of a single atom. The optical pumping laser for the two atoms differs by a relative phase of $\pi$.
         (b) The effective transitions for two-atom trapped system. With the effective driving fields and decays, ultimately, the system will be stabilized into the state $|S\rangle$.
         }
 \label{fig21}
\end{figure}

\textit{Application II: two-atom entanglement.} We consider two $\Lambda$ atoms with a level structure, as shown in Fig. \ref{fig21} (a)
(marked as atom $A$ and atom $B$), which are trapped in an optical cavity.
The transition $|g_{1}\rangle_{m}\leftrightarrow|e\rangle_{m}$ ($m=A,B$) is resonantly driven by a laser with Rabi frequency $\Omega_{m}$,
and the transition $|g_{2}\rangle_{m}\leftrightarrow|e\rangle_{m}$ is coupled to the quantized cavity field resonantly with
coupling strength $\lambda$. Besides, we apply a microwave field with Rabi frequency $\Omega_{MW}$ to drive
the transition between ground states $|g_{1}\rangle_{m}$ and $|g_{2}\rangle_{m}$ with detuning $\delta$.
The Hamiltonian for this system in an interaction picture reads
\begin{small}
\begin{align}\label{eq2-1}
  H_{0}=&
  \sum_{m=A,B}\Omega_{m}|e\rangle_{m}\langle g_{1}|+\Omega_{MW}e^{i\delta t}|g_{2}\rangle_{m}\langle g_{1}| \cr
         &+\lambda|e\rangle_{m}\langle g_{2}|a+H.c., 
\end{align}
\end{small}
where $a$ denotes the cavity annihilation operator.
The corresponding dynamics of the current system is described by the master equation in Eq. (\ref{eq0-1}).
The Lindbald operators associated with atomic
spontaneous emission and cavity decay are
  $L_{m_{1}}=\sqrt{\gamma_{1}/2}|g_{1}\rangle_{m}\langle e|$, $L_{m_{2}}=\sqrt{\gamma_{2}/2}|g_{2}\rangle_{m}\langle e|$ ($m=A,B$),
  and $L_{C}=\sqrt{\kappa}a=\sqrt{\kappa}|0\rangle_{C}\langle 1|$,
where $\kappa$ is the cavity decay rate and $|k\rangle_{C}$ ($k=0,1$) denotes the photon number in the cavity.

Referring to the formula of quantum Zeno dynamics \cite{Prl89080401Jpa41493001}, we write the Hamiltonian $H_{0}$ as
$H_{0}=\Omega(H_{p}+K H_{q})$, where $\Omega=\sqrt{\Omega_{A}^2+\Omega_{B}^2+\Omega_{MW}^{2}}$, $K=g/\Omega$,
$H_{p}$ stands for the dimensionless interaction Hamiltonian between the atom and the
classical field, and $H_{q}$ denotes the counterpart between the atom
and the quantum cavity field. When the strong coupling limit $K\rightarrow\infty$ is satisfied,
we obtain the effective Hamiltonian $H_{0}^{eff}=\Omega_{a}(\sum_{l}P_{l}H_{p}P_{l}+K\epsilon_{l}P_{l})$,
where $P_{l}$ is the eigenprojection and $\epsilon_{l}$ is the corresponding
eigenvalue of $H_{q}$: $H_{q}=\sum_{l}\epsilon_{l} P_{l}$.
Assuming the system is initially in the Zeno dark subspace ($\epsilon_{l}=0$) spanned by
  $|\psi_{1}\rangle=|g_{1}g_{2}\rangle_{A,B}|0\rangle_{C}$, $|\psi_{2}\rangle=|g_{2}g_{1}\rangle_{A,B}|0\rangle_{C}$,
  $|\psi_{3}\rangle=|g_{2}g_{2}\rangle_{A,B}|0\rangle_{C}$, $|\psi_{4}\rangle=|g_{1}g_{1}\rangle_{A,B}|0\rangle_{C}$,
  and $|D\rangle=\frac{1}{\sqrt{2}}(|eg_{2}\rangle_{A,B}-|g_{2}e\rangle_{A,B})|0\rangle_{C}$,
the effective Hamiltonian reduces to ($\Omega_{A}=-\Omega_{B}$ and $\Omega_{0}=\sqrt{\Omega_{A}^2+\Omega_{B}^2}$)
\begin{small}
\begin{align}\label{eq2-4}
  H_{0}^{eff}=&\frac{\Omega_{0}}{\sqrt{2}}|D\rangle\langle T|+\sqrt{2}\Omega_{MW}e^{i\delta t}|\psi_3\rangle\langle T| \cr
              &+\sqrt{2}\Omega_{MW}e^{-i\delta t}|\psi_{4}\rangle\langle T|  +H.c.,
\end{align}
\end{small}
where
$|S\rangle=(|\psi_{1}\rangle-|\psi_{2}\rangle)\sqrt{2}$ and
$|T\rangle=(|\psi_{1}\rangle+|\psi_{2}\rangle)\sqrt{2}$.
Accordingly, the effective Lindblad operators are
  $\tilde{L}_{G}=\sqrt{{\gamma_{2}}/{2}}|\psi_{3}\rangle\langle D|$,
  $\tilde{L}_{S}=\sqrt{{\gamma_{1}}/{4}}|S\rangle\langle D|$,
  and $\tilde{L}_{T}=\sqrt{{\gamma_{1}}/{4}}|T\rangle\langle D|$.
The cavity field has been decoupled
in the effective Hamiltonian when the Zeno condition is satisfied thus the cavity
decay can be neglected.
Figure \ref{fig21} (b) shows the effective transitions of reduced system.

\begin{figure}[t]
  \centering
 \scalebox{0.32}{\includegraphics {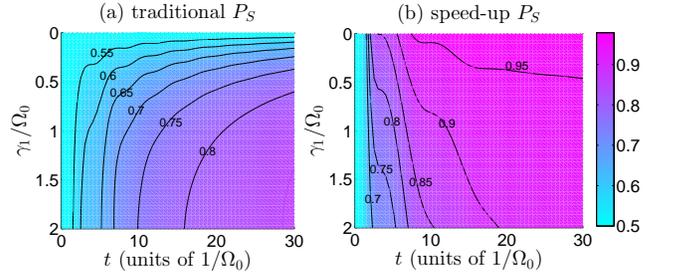}}
 \caption{
         Two-atom entanglement preparation:
         The comparison with respect to the two-atom entanglement generation between the traditional dissipation dynamics and
         the accelerated dissipation dynamics.
         (a) $P_{S}$ versus $\gamma_{1}$ via traditional dissipation dynamics.
         (b) $P_{S}$ versus $\gamma_{1}$ via accelerated dissipation dynamics.
         The basic parameters in plotting the figure are $\lambda=10\Omega_{0}$, $\delta=0.15\Omega_{0}$,
         $\kappa=0.5\Omega_{0}$, $\gamma_{2}=0.5\gamma_{1}$, and $\Omega_{MW}=0.2\Omega_{0}$. The initial state is selected as $\rho_{0}=|\psi_{1}\rangle\langle\psi_{1}|$.
         }
 \label{fig22}
\end{figure}

\begin{figure}[t]
  \centering
 \scalebox{0.6}{\includegraphics {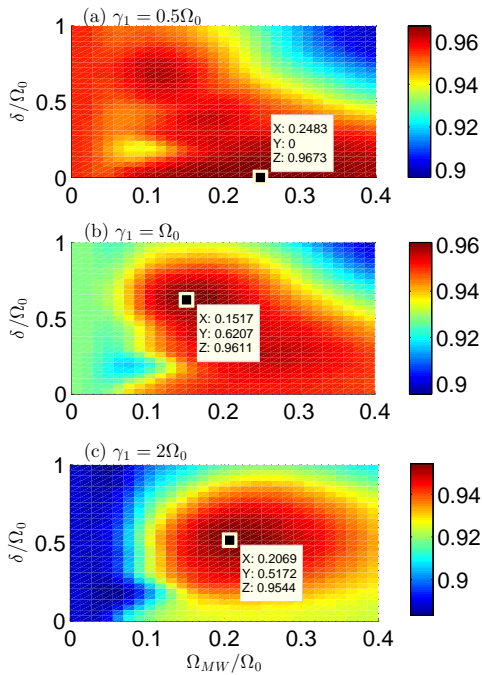}}
 \caption{
         Two-atom entanglement preparation:
         The fidelity $F_{S}$ versus detuning $\delta$ and Rabi frequency $\Omega_{MW}$ with
         (a) $\gamma_{1}=0.5\Omega_{0}$;
         (b) $\gamma_{1}=\Omega_{0}$;
         (c) $\gamma_{1}=2\Omega_{0}$.
         The basic parameters in plotting the figure are $\lambda=10\Omega_{0}$,
         $\kappa=0.5\Omega_{0}$, and $\gamma_{2}=0.5\gamma_{1}$. The initial state is selected as $\rho_{0}=|\psi_{1}\rangle\langle\psi_{1}|$.
         }
 \label{fig6}
\end{figure}

The time-dependent population
for the target state $|S\rangle$ versus decay rate $\gamma_{1}$ is shown in Fig. \ref{fig22} (a).
Obviously, an operation time $\mathcal{T}_{t}=30/\Omega_{0}=300/\lambda$ is not enough to
generate the entangled state $|S\rangle$
[the maximum population for $|S\rangle$ in Fig. \ref{fig5} (a) is only $0.8548$].
A further study shows that for $\gamma_{1}\leq2\Omega_{0}$, an operation time $\mathcal{T}_{t}\geq 1000/\lambda=100/\Omega_{0}$
is necessary in order to obtain a relatively high-fidelity ($F_{S}\geq0.9$) entanglement.
Such results can be also found in the previous schemes for the generation of two-atom entanglement.
For example, in Ref. \cite{Prl106090502}, by choosing parameters similar to those in plotting Fig. \ref{fig22} (a),
the time required for entanglement generation with fidelity $F_{S}\geq 0.9$
is $\mathcal{T}_{t}\geq 1300/\lambda=130/\Omega_{0}$.
The control Hamiltonians to accelerate entanglement generation are chosen as
  $H_{1}=\mu_{1}|e\rangle_{A}\langle g_{1}|+H.c.$ and
  $H_{2}=\mu_{2}|e\rangle_{B}\langle g_{1}|+H.c.$.
We randomly select $\mu_{1}=1$ and $\mu_{2}=1.5$ as an example
to show time-dependent $P_{S}$ versus $\gamma_{1}$ in
Fig. \ref{fig22} (b).
One can find from Fig. \ref{fig22}
that the entanglement generation has been accelerated by the additional coherent control fields.
An operation time $\mathcal{T}_{a}\leq 20/\Omega_{0}$ is enough to generate two-atom entanglement
with fidelity $F_{S}\geq0.9$. In fact, by choosing suitable parameters for a specified decay rate, the fidelity can be further improved (See Fig. \ref{fig6}).
As shown in the figure, for decay rate $\gamma_{1}=0.5\Omega_{0}$ [See Fig. \ref{fig6} (a)],
the optimal parameters are $\delta=0$ and $\Omega_{MW}\sim0.25\Omega_{0}$, and the corresponding fidelity is $F_{S}\sim0.97$;
for decay rate $\gamma_{1}=\Omega_{0}$ [See Fig. \ref{fig6} (b)],
when $\delta\sim0.6\Omega_{0}$ and $\Omega_{MW}\sim0.15\Omega_{0}$, we have the highest fidelity $F_{S}\sim0.96$;
for decay rate $\gamma_{1}=2\Omega_{0}$ [See Fig. \ref{fig6} (c)],
the highest fidelity $F_{S}\sim0.96$ appears when $\delta\sim0.5\Omega_{0}$ and $\Omega_{MW}\sim0.2\Omega_{0}$.
The experimentally achievable values for cooperativity are around
$C=\lambda^2/(\gamma_{1}\kappa)\approx 100$ \cite{Prl97083602Prl101203602},
corresponding to $\gamma_{1}\approx2\Omega_{0}$ and $\kappa\approx0.5\Omega_{0}$. 
For $\lambda=(2\pi)35$MHz, with the experimentally achievable parameters,
the operation time required for the entanglement generation is only about 1.3 $\mu$s,
which is much shorter than the typical decoherence time scales for this system.

In conclusion, we have investigated the possibility of accelerating dissipation-based
state generation in a three-level system and a trapped two-atom
system. From both analytical and numerical evidence, we have shown that the speed for a system to reach the
target state has been significantly improved with additional coherent control fields,
without losing the advantage of robustness against parameter fluctuations.
Notably, the additional control fields are given basically according to the definition of the system evolution speed
via dissipation dynamics [see Eq. (\ref{eq0-4})], while there are in fact other definitions that can be used and the
control fields would be accordingly changed. So, in the future, it would be interesting to study
the behavior of the given additional coherent control fields based on other definitions of the evolution speed.

This work was supported by the National Natural Science Foundation
of China under Grants No. 11575045, No. 11374054 and No. 11674060.

\end{document}